\documentclass[aps,pra,twocolumn,showpacs,superscriptaddress]{revtex4}

\usepackage{epsfig}
\usepackage{graphicx}
\usepackage{enumitem}

\begin{document}

\title{First Long-Term Application of Squeezed States of Light in a
Gravitational-Wave Observatory}

\author{H. Grote}
\email{hartmut.grote@aei.mpg.de}
\author{K. Danzmann}
\author{K.L. Dooley}
\author{R. Schnabel}
\author{J. Slutsky}
\author{H. Vahlbruch}
\affiliation{Max-Planck-Institut f\"ur Gravitationsphysik (Albert Einstein Institut) und
Leibniz Universit\"at Hannover, Callinstr.\,38, 30167 Hannover, Germany}

\raggedbottom

\date{\today}

\begin{abstract}
We report on the first long-term application of squeezed vacuum states
of light to improve the shot-noise-limited sensitivity of a
gravitational-wave observatory. In particular, squeezed vacuum was
applied to the German\,/\,British detector GEO\,600 during a period of
three months from June to August 2011, when GEO\,600 was performing an
observational run together with the French\,/\,Italian Virgo detector.
In a second period squeezing application continued for about 11 months
from November 2011 to October 2012.  During this time, squeezed vacuum
was applied for 90.2\% (205.2 days total) of the time that
science-quality data was acquired with GEO\,600.
Sensitivity increase from squeezed vacuum application was observed
broad-band above 400\,Hz.  The time average of gain in sensitivity was
26\,\% (2.0\,dB), determined in the frequency band from 3.7\,kHz to
4.0\,kHz.  This corresponds to a factor of two increase in observed
volume of the universe, for sources in the kHz region
(e.g. supernovae, magnetars).
We introduce three new techniques to enable stable long-term
application of squeezed light, and show that the glitch-rate of the
detector did not increase from squeezing application.
Squeezed vacuum states of light have arrived as a permanent
application, capable of increasing the astrophysical reach of
gravitational-wave detectors.
\end{abstract}
  

\maketitle


Gravitational waves were predicted by Albert Einstein as a consequence
of his General Theory of Relativity \cite{Ein16}, but have not been
directly measured to date.  While a number of large-scale
laser-interferometers such as LIGO~\cite{AbbottETAL09ligo},
Virgo~\cite{Accadia12}, TAMA~\cite{AraiETAL09}, and
GEO\,600~\cite{Wetal02}, were (or are, in the case of GEO\,600)
in operation, their sensitivities were not yet sufficient to make the
first direct detection of gravitational waves (GW).  One of the
sensitivity limits of these instruments originates from the zero-point
fluctuation of the quantum vacuum state of light.  While this noise
can in principle be reduced with respect to the signal size by the use
of higher laser power, this method is ultimately limited by thermal
deformation of optical components and other laser power related
instabilities~\cite{Dambr03,Zhao05}.  It was proposed by
Caves~\cite{Cav81} that quantum noise in the shot-noise regime can
also be reduced with the injection of squeezed vacuum states into the
otherwise open output port of an interferometer.  It took several
years until squeezed-light sources became available and to demonstrate
this principle on table-top
interferometers~\cite{XIAO87,GRAN87,MSMBL02,VCHFDS05}.  Even more time
was required to develop squeezed-light sources that made squeezed
states at audio frequencies available, which is the frequency band
relevant for the km-scale
interferometers~\cite{Metal04,VCHFDS06,VCDS07,VMCH08}.  Squeezing
above the audio band was demonstrated in a prototype
GW-interferometer~\cite{Getal08} with suspended mirrors, and the first
demonstration of quantum-enhancement in one of the gravitational-wave
observatories was done in 2010 in the GEO\,600
detector~\cite{Schnabel2011}. Squeezing of the LIGO observatory in
Hanford, WA, was then demonstrated in late 2011~\cite{TBP}.  While all
of these experiments were one-time demonstrations of squeezed vacuum
applied on time scales of minutes, in this work we report on the first
continuous (and ongoing) application of squeezed light to a
gravitational-wave interferometer, namely GEO\,600,
to enhance the shot-noise limited sensitivity.



The GEO detector is currently the only operational
laser-interferometric GW detector worldwide and is acquiring
calibrated measurement data in the so-called \emph{astrowatch}
mode. Data taken in this mode covers about 2/3 of the total time, when
regular work as part of an incremental upgrade program 
called \emph{GEO-HF}~\cite{Luetal10}
does not take place. The search for gravitational-wave
signals in the astrowatch data will be triggered only by the
occurrence of external (e.g. electromagnetic or neutrino) signals, as
expected e.g. from a supernova in our galaxy.

\begin{figure*}[t]
\includegraphics[width=15cm]{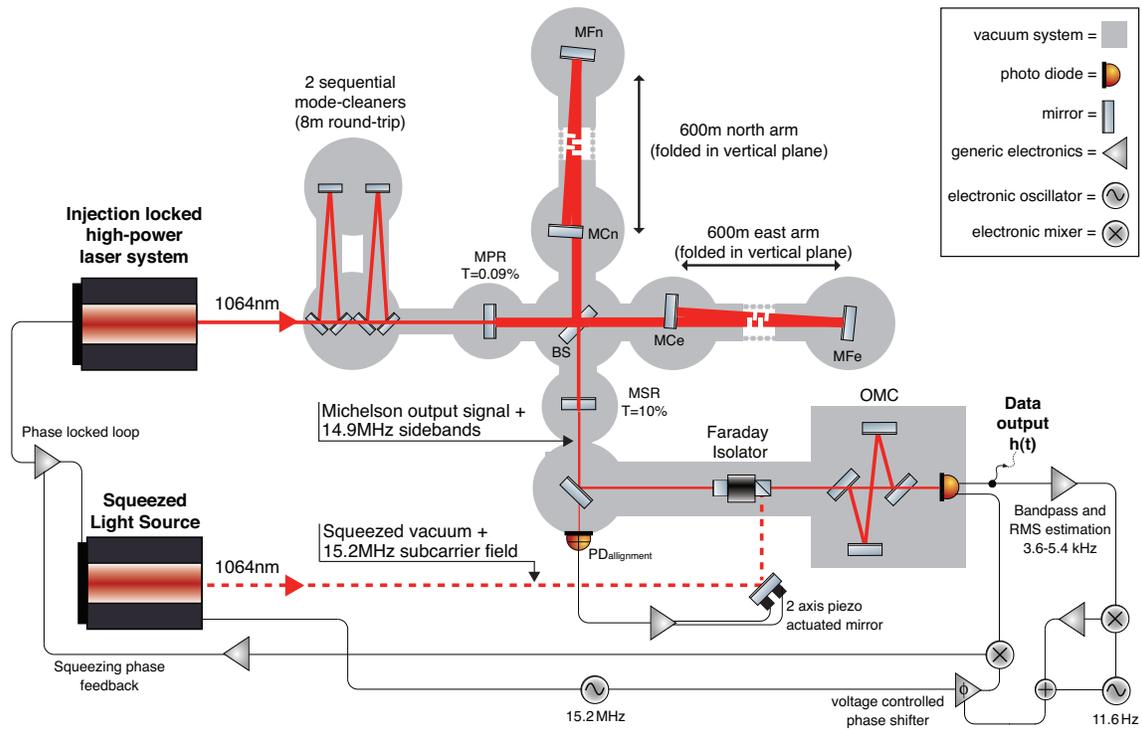}
\vspace{0mm}
\caption{A simplified optical layout of the squeezed-light enhanced
  gravitational wave detector GEO\,600, which consists of the
  conventional GEO\,600 observatory and the additional squeezed-light
  source. The observatory has two singly folded arms with a total
  optical length of 2400\,m. A gravitational wave passing from most
  directions will shorten one arm, while the length of the
  perpendicularly orientated arm is increased, and vice versa in the
  next half-cycle of a passing wave, producing a periodic power change
  of the output light that is detected by a photo diode. The
  observatory is operated such that almost of all the light is
  back-reflected towards the input laser by keeping the interferometer
  output close to a dark fringe via a control system. A
  power-recycling mirror (MPR) leads to a resonant enhancement of the
  circulating light power of 2.7\,kW at the beam splitter.  Similar to
  the power-recycling technique, a partially transmissive
  signal-recycling mirror (MSR) is installed to further resonantly
  enhance the GW-induced signal at the interferometer's output.  BS:
  50/50 beam splitter, MFe/MFn: far interferometer end mirrors
  (east/north), MCe/MCn: central interferometer mirrors, T: mirror
  power transmissivity. All interferometer optics are situated in a
  vacuum system and suspended by multi-stage pendulums.}
\label{GEOsetup}
\end{figure*}

\begin{figure}[Htb]
\centering
\includegraphics[width=1.0\linewidth]{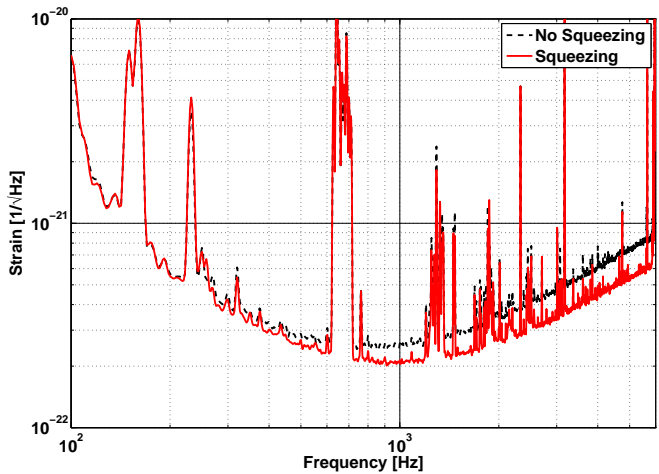}
\caption{Strain amplitude spectral densities of GEO\,600 with and
  without squeezing being applied. This measurement reflects the state
  of squeezing application at the end of August 2012, as an example
  for the squeezing performance during the time from November 2011 to
  October 2012.  The large structures from 620\,--\,720\,Hz are
  unresolved lines caused by vibrational modes of the wires suspending
  the test masses.}
\label{strain_fig}
\end{figure}

GEO\,600~\cite{Wetal02,Gretal10} is a Michelson interferometer with
folded 600\,m long arms, using the advanced technique of dual
recycling~\cite{Meers88} (power- and signal recycling). Figure
\ref{GEOsetup} shows a simplified layout of GEO\,600, together with
the squeezed-light source ~\cite{VKLGDS10,KVLGDGS12}. The squeezed
states are injected into the main interferometer via a Faraday
Isolator and must be mode-matched and aligned to the main
interferometer beam. In addition, two servo systems (denoted in
Fig.~\ref{GEOsetup} as `Phase locked loop' and `Squeezing phase
feedback') are necessary to assure the squeezing is at the correct
optical laser frequency and detected interferometer output
quadrature. Through the implementation of three new techniques for
alignment and squeezing phase control, we achieve a stable and
long-term application of squeezed light to a GW
observatory. Figure~\ref{strain_fig} shows typical strain spectral
densities of GEO\,600, comparing the non-squeezed noise floor with the
squeezed noise floor, at the end of August 2012. Shot-noise reduction
is observed at frequencies higher than about 400\,Hz.

We apply an automatic alignment system for a squeezed vacuum field for
the first time. Because all of the GEO\,600 core optics are suspended
as pendula to reduce seismic perturbations, the alignment of the
squeezed field into the interferometer must continuously follow the
spatially drifting GEO\,600 output field.
An error signal for this control loop is derived by sensing the
optical beat between the 15.2\,MHz frequency shifted squeezed vacuum
control field~\cite{VCHFDS06} and the 14.9\,MHz Michelson
interferometer sidebands leaving the interferometer's dark
port~\cite{Gretal10}, as illustrated in Fig.~\ref{GEOsetup}. The servo
feedback signals are sent to a 2-axis piezo actuated mirror, placed in
the squeezing injection path, suppressing relative motions up to a
frequency of 3\,Hz. 

Furthermore, we introduce a new control signal for the phase relation
between the injected squeezed field and the interferometer output
field. We derive an error signal using the detected light transmitted
through the output mode-cleaner (OMC) where the squeezing coherent
control sidebands~\cite{VCHFDS06} beat with the spatially filtered
GEO\,600 carrier light (TEM00-mode).
The reduction of higher order modes in the filtered GEO output beam,
as well as the fact that all of the output TEM00-mode carrier light is
available to use, results in a superior error signal to shot noise
ratio, as compared to previously used techniques. Although the
detected squeezing level is currently primarily limited by optical
loss~\cite{Schnabel2011}, employing this new control signal already
improves the strain sensitivity of GEO 600. As optical losses are
reduced, which is foreseen during the GEO-HF upgrade, phase noise will
play a more significant role in limiting the detected squeezing level,
thus making the benefits of this technique more important. The new
control signal will also be of particular interest to the third
generation laser interferometeric GW-detectors~\cite{Hildal11} where
the goal is to detect up to 10\,dB of squeezing.


Finally, the coherent control scheme for controlling the squeezer
phase is extended to include a noise locking
scheme~\cite{McKenzie05} to compensate for slow drifts of the 
optimal squeezing angle. The implementation of this noise lock
is similar to the one described in \cite{McKenzie05}, but the combination
with the coherent control scheme is used for the first time.
The demodulation phase of the squeezing phase control signal is
electronically modulated at a frequency of 11.6\,Hz. In order to
maximize the detected squeezing level, as measured in a frequency band
between 3.6\,kHz and 5.4\,kHz of the h(t) data stream, a control loop
with unity gain at about 10\,mHz corrects for drifts of the optimal
demodulation phase over long timescales. 
The unity gain frequency of the coherent control loop is about 1\,kHz.

The three new techniques described above were implemented in stages
during two long-term GEO\,600 science data taking periods. From June
3 to September 5, 2011, GEO\,600 took part in an observational
run, named S6e/VSR4, together with the Virgo detector.  During the
93.3 days of this run, no detector commissioning work was performed.
As a result, science-quality data were acquired for 84.3\,\% (78.8
days) of the total time, with squeezing applied for 53.1\,\% (41.8
days) of the science time. The average squeezing level observed in
this period was 16\,\% (1.3\,dB).

In the second period lasting from November 17, 2011 to October 15,
2012, the GEO\,600 observatory took data in regular astrowatch mode.
Compared to the S6e/VSR4 run, we achieved a much higher squeezing
application time, as well as an improved observed average squeezing
level.  The astrowatch time obtained in this period of 333 days was
68.2\,\% (227.4 days) with squeezed vacuum being applied for 90.2\,\%
(205.2 days) of the astrowatch time with an average gain in detector
sensitivity of 26\,\% (2.0\,dB). For the last two weeks of this period
we achieved the highest duty factor so far: squeezing was applied to
99.66\,\% of the astrowatch science-quality data, demonstrating the
success of the new control techniques.


\begin{figure}[Htb]
\centerline{
\includegraphics[width=1.0\linewidth]{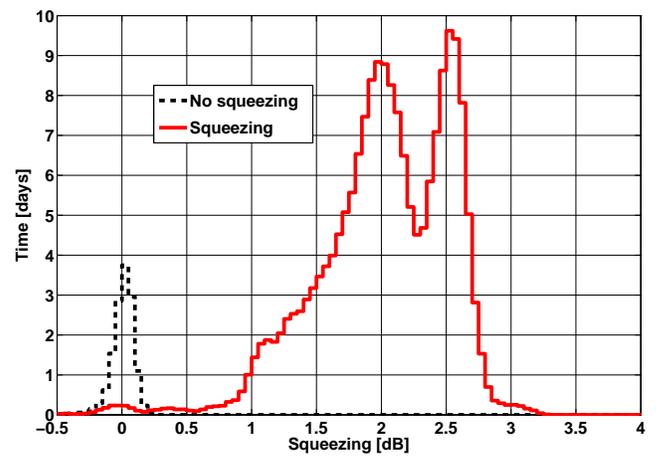}}
\vspace{0mm}
\caption{Histogram of the observed squeezing at GEO\,600 in the band
  from 3692 to 3957\,Hz. A median of 26\,\% (2.0\,dB) of squeezing was
  achieved when squeezing was applied. This histogram contains the
  full data set from November 17, 2011 to October 15, 2012. The
  different squeezing levels are caused mainly by different operating
  conditions of the squeezing application. The larger squeezing level
  of about 2.5\,dB is mainly from the latest time in the period
  reported here when all new control techniques were being used.}
\label{hist_fig}
\end{figure}

In order to characterise the effect of the squeezing application we look 
at the frequency dependent increase in sensitivity, as well as the 
stationarity of the data.  We use the Omega data analysis 
pipeline~\cite{Chatterji:2004qg} for these investigations, which calculates the 
amplitude spectral density of 64-second data intervals, overlapping by 8 
seconds, starting from the beginning of each science data segment. For the 
squeezing level estimation we compare this amplitude spectrum with the 
expected un-squeezed shot-noise amplitude spectrum, evaluating for the 
frequency range from 3692\,Hz to 3957\,Hz. 
This range is chosen as it is 
clean of spectral lines and other noise contributions, simplifying the 
long term calculation of the squeezing. However, the suppression is broadly 
the same in the 2\,kHz to 6\,kHz regime (see also Figure~\ref{strain_fig}). 
Figure~\ref{hist_fig} shows the result for the period from November 17, 
2011 to October 15, 2012.  The average gain in sensitivity from squeezed 
vacuum application in this period was 26\,\% (2.0\,dB). This improvement in 
sensitivity corresponds to an increase of the observable volume of the 
universe by a factor of two for gravitational-wave sources in this 
frequency band.

To investigate the noise behavior on short time scales (i.e. 1 second and 
below) the Omega pipeline uses the 64\,s long spectra to whiten the data, 
and then performs a wavelet analysis to search for transients, which are 
thresholded and clustered into triggers. These transients, named glitches 
for our purpose, degrade our ability to detect potential 
gravitational-wave signals with similar time-frequency parameters. 
Glitches occur in all gravitational-wave detectors and can have many 
different origins (i.e. optical, electrical, etc., see e.g.~\cite{Aasi12}).
Any new technique added to a gravitational-wave detector 
should not increase the glitch rate, which would compromise the potential 
utility of the technology.

\begin{figure}[Htb]
\centering
\includegraphics[width=1.0\linewidth]{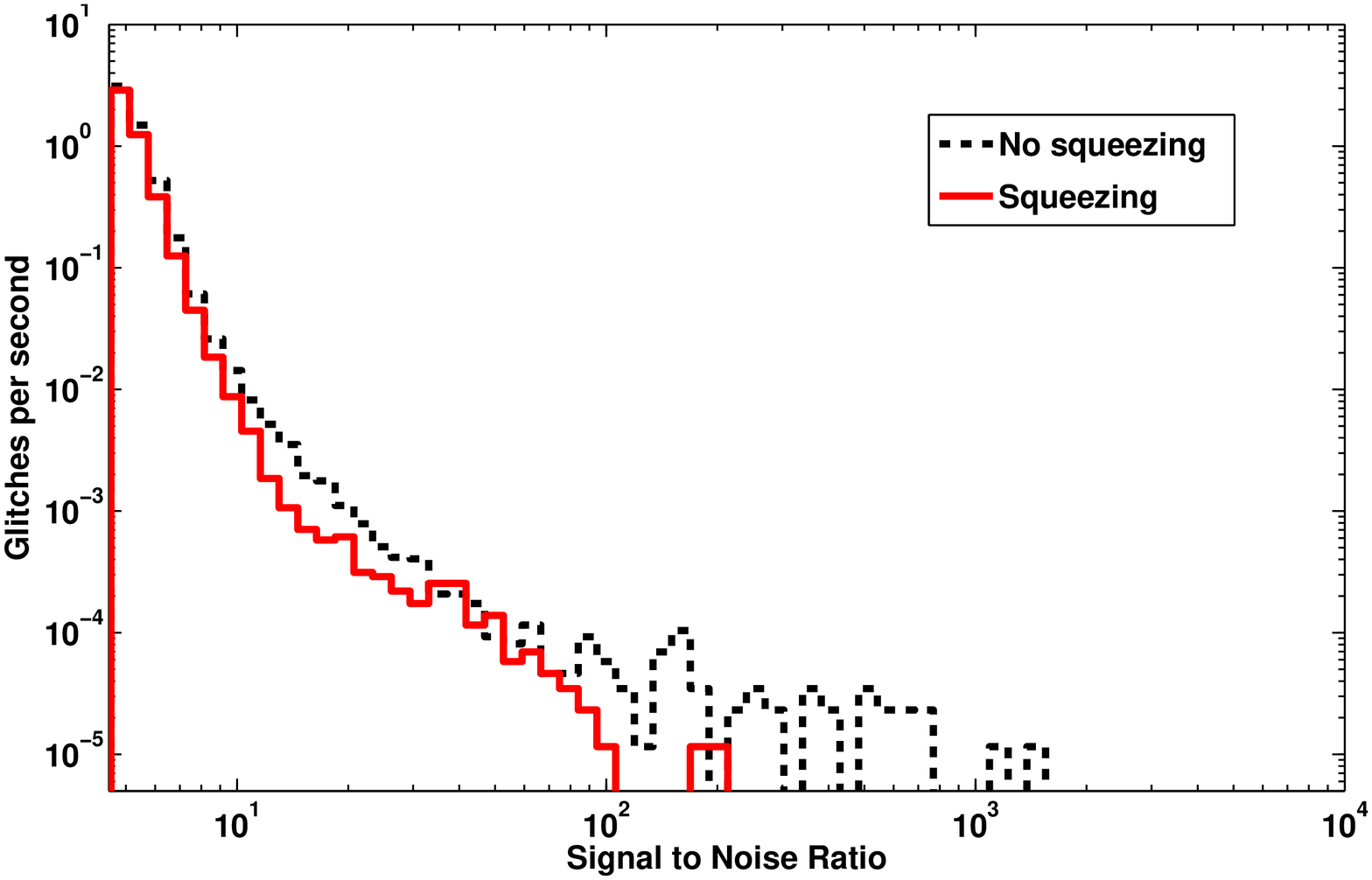}
\caption{Histograms of the signal-to-noise ratio (SNR) of glitches in the
gravitational-wave channel. Two data sets of 24 hours duration each 
are compared
for the cases of squeezing being applied and not being applied,
respectively. No increase in glitch rate can be observed 
for the case of squeezing being applied. The reduction in glitch rate
when squeezing was applied is not considered significant with respect to the
usage of squeezing. 
}
\label{glitch_fig}
\end{figure}

Figure \ref{glitch_fig} shows histograms of signal-to-noise ratios for
glitches detected during two 24 hour reference periods, one each for
squeezing and no squeezing applied, constituting the first
investigation of this kind. Neither the rate nor outlier amplitude of
the glitches was observed to increase for the interval with squeezing
applied, compared with the time without the application of squeezing.
This data without the application of squeezing was created
intentionally to examine the effect of the squeezer on the glitch
population in the detector data.  No other deliberate attempt was made
to set aside data without squeezing, but investigation of a similar
data stretch from months earlier when squeezing was unavailable for
technical reasons yielded consistent results.

We will continue to apply squeezed vacuum as much as
possible during the next years, while also trying to increase
the level of observed squeezing. Currently, the main limitation 
in the level of observed squeezing comes from optical losses
in the output-mode cleaner.

We conclude that we have demonstrated the application of squeezed vacuum 
states over a period of more than one year, enhancing the shot-noise limited 
sensitivity of a laser-interferometric gravitational-wave observatory. 
Squeezed vacuum states are now regularly used at GEO\,600.

\section*{Acknowledgments}
We thank the GEO collaboration for the construction of GEO\,600 
and the quantum interferometry group for the development of the squeezed light 
source and its technology.
The authors are also grateful for support from 
the Science and Tech. Facilities Council (STFC), 
the Univ. of Glasgow in the UK, 
the Max Planck Society,  
the Bundesm. f. Bildung und Forschung (BMBF), 
the Volkswagen Stiftung, 
the cluster of excellence QUEST, 
the Max Planck Research School (IMPRS),
the State of Nds. in Germany,
and DFG grant SFB/TR 7.



\begin{thebibliography}{3}

\bibitem{Ein16} Einstein,A. 
\textit{Annalen der Physik} \textbf{49}, 769-822 (1916).

\bibitem{AbbottETAL09ligo} Abbott,B.~P.~et al. 
\textit{Rep. Prog. Phys.} \textbf{72}, 076901 (2009).

\bibitem{Accadia12} Accadia,T.~et al. 
\textit{Journal of Instrumentation} \textbf{7}, P03012 (2012).

\bibitem{AraiETAL09} Arai,K.~et al. 
\textit{Class. Quantum Grav.} \textbf{26}, 204020 (2009).

\bibitem{Wetal02} Willke,B.~et al. 
\textit{Class. Quantum Grav.} \textbf{19}, 1377-1387 (2002).

\bibitem{Dambr03} DAmbrosio,E.
\textit{Phys. Rev. D} \textbf{67}, 102004 (2003). 

\bibitem{Zhao05} Zhao,C. Ju,L. Degallaix,J. Gras,S. Blair,D.G.
\textit{Phys. Rev. Lett.} \textbf{94} 121102 (2005) 

\bibitem{Cav81} Caves,C.~M. 
\textit{Phys. Rev. D} \textbf{23}, 1693-1708 (1981).

\bibitem{XIAO87} Xiao,M. Wu,L.A.~\&~Kimble, H. J. 
\textit{Phys. Rev. Lett.} \textbf{59}, 278-281 (1987).

\bibitem{GRAN87} Grangier,P. Slusher,R.E., Yurke,B.~\&~LaPorta,A.
\textit{Phys. Rev. Lett.} \textbf{59}, 2153-2156 (1987).

\bibitem{MSMBL02} McKenzie,K. Shaddock,D.~A. McClelland,D.~E. Buchler,
 B.~C.~\&~Lam,P.~K. 
\textit{Phys. Rev. Lett.} \textbf{88}, 231102 (2002).

\bibitem{VCHFDS05} Vahlbruch,H. Chelkowski,S. Hage,B. Franzen,A. Danzmann,K.~\&~Schnabel,R. \textit{Phys. Rev. Lett.} \textbf{95}, 211102 (2005).

\bibitem{Metal04} McKenzie,K.~et al. 
\textit{Phys. Rev. Lett.} \textbf{93}, 161105 (2004).

\bibitem{VCHFDS06} Vahlbruch,H. Chelkowski,S. Hage, B. Franzen,A.
  Danzmann,K.~\&~Schnabel,R. 
\textit{Phys. Rev. Lett.} \textbf{97}, 011101 (2006).

\bibitem{VCDS07} Vahlbruch,H. Chelkowski,S. Danzmann,K. Schnabel,R.
\textit{New J. Phys.} \textbf{9}, 371 (2007).

\bibitem{VMCH08} Vahlbruch,H.~et al.
\textit{Phys. Rev. Lett.} \textbf{100}, 033602 (2008). 

\bibitem{Getal08} Goda,K.~et al. 
\textit{Nature Physics} \textbf{4}, 472-476 (2008).

\bibitem{Schnabel2011}
The LIGO Scientific Collaboration,
\textit{Nature Physics} \textbf{7}, 962-965 (2011)

\bibitem{TBP} The LIGO Scientific Collaboration
%
\textit{to be published} 

\bibitem{Luetal10} L\"uck, H.~et al. 
\textit{J. Phys. Conf. Ser.} \textbf{228}, 012012 (2010).


\bibitem{Gretal10} Grote,H.~et al. 
\textit{Class. Quantum Grav.} \textbf{27}, 084003 (2010).

\bibitem{Meers88} Meers,B.J. 
\textit{Phys. Rev.} \textbf{D 38}, 2317-2326 (1988).

\bibitem{VKLGDS10} Vahlbruch,H. Khalaidovski,A. Lastzka,N. Gr{\"a}f,C.
  Danzmann,K.~\&~Schnabel,R. 
{\em Class. Quantum Grav.} {\bf 27}, 084027 (2010).

\bibitem{KVLGDGS12} Khalaidovski,A. Vahlbruch,H. Lastzka,N. Gr{\"a}f,C.
  Danzmann,K. Grote,H.~\&~Schnabel,R. 
{\em Class. Quantum Grav.} {\bf 29}, 075001 (2012).


\bibitem{Hildal11} Hild, S.~et al. 
\textit{Class. Quantum Grav.} \textbf{28}, 094013 (2011).

\bibitem{McKenzie05} McKenzie, K.~et al. 
\textit{J. Opt. B: Quant. Semiclass. Opt.} \textbf{7}, 421-428 (2005).


\bibitem{Chatterji:2004qg} Chatterji,S. Blackburn,L. Martin,G. and
  Katsavounidis,E.
{\em Class. Quant. Grav.} {\bf 21} 1809-1818 (2004)

\bibitem{Aasi12} Aasi,J.~et al. 
{\em Class. Quantum Grav.} {\bf 29} 155002 (2012)



\end{thebibliography}
\end{document}